\documentclass[aip,rsi,reprint,graphicx,draft]{revtex4-1} 

\usepackage[final]{graphicx}
\graphicspath{{figures/}}
\usepackage[caption=false]{subfig}
\usepackage{siunitx}
\DeclareSIUnit{\dBm}{dBm}
\usepackage{xspace}
\usepackage{amsmath}
\newcommand{\RFM}{\SI{2856}{\MHz}\xspace}
\newcommand{\RFG}{\SI{2.856}{\GHz}\xspace}

\begin{document}


\title{Electron bunch energy and phase feed-forward stabilization system for the Mark~V RF-linac free-electron laser} 



\author{M.R. Hadmack}
\affiliation{Department of Physics and Astronomy, University of Hawai`i at M\=anoa, Honolulu, Hawai`i, USA 96822}
\author{B.T. Jacobson}
\affiliation{RadiaBeam Technologies, Santa Monica, California, USA 90404}
\author{J.M.D. Kowalczyk}
\author{B.R. Lienert}
\author{J.M.J. Madey}
\author{E.B. Szarmes}
\affiliation{Department of Physics and Astronomy, University of Hawai`i at M\=anoa, Honolulu, Hawai`i, USA 96822}



\date{\today}

\begin{abstract}
An amplitude and phase compensation system has been developed and tested at the University of Hawai`i for the optimization of the RF drive system to the Mark~V Free-Electron Laser.  
Temporal uniformity of the RF drive is essential to the generation of an electron beam suitable for optimal free-electron laser performance and the operation of an inverse Compton scattering x-ray source.  
The design of the RF measurement and compensation system is described in detail and the results of RF phase compensation are presented.   
Performance of the free-electron laser was evaluated by comparing the measured effects of phase compensation with the results of a computer simulation.
Finally, preliminary results are presented for the effects of amplitude compensation on the performance of the complete system.
\end{abstract}


\maketitle 

\section{Introduction}
A critical factor in the performance of RF linac free-electron lasers (FELs) is the amplitude and phase uniformity of the high-power microwave pulses used to drive the electron gun and linac.  
Amplitude non-uniformity contributes to an energy spread among the electron bunches and a loss of laser efficiency due to frequency pulling.  
Phase irregularities in the RF waveform result in variations in the arrival time of the electron bunches, which reduces laser gain due to a loss of synchronism between the electron bunches and the circulating optical pulses in the laser.
For example, a phase excursion of \ang{1} at \RFM delays the electron bunches by \SI{1}{\ps}, which is the full width of a typical optical pulse, leading to reduced laser gain and reduced laser power.

Electron bunch synchronization is also of particular importance for the operation of optical storage cavity-based Compton x-ray sources, such as the system currently under development at the University of Hawai`i\cite{hadmack2012phd}.
Incoming electron bunches must be synchronized to the circulation time of laser pulses in the storage cavity, and any loss of beam synchronism in the scattering region will lead to reduced x-ray flux.

A high power ITT Triton 2960 klystron is used to power the MkV FEL's \SI{40}{\MeV} Linac; it amplifies low phase noise, low-level microwave pulses to the \SI{30}{\mega\watt} power level required for the acceleration of a relativistic electron beam.  
The accelerating potential of the klystron amplifier is responsible for the device's gain and takes the form of a \SI{300}{\kV} pulse.  
Uniformity of the klystron gain requires careful regulation of the pulse amplitude over its \SI{\sim 8}{\us} duration.  
This pulse results in a high voltage DC potential applied to the klystron, discharged through activation of a hydrogen thyratron switch via an LC resonant pulse forming network (PFN).  
Tuning the PFN resonant circuit allows the high voltage impulse through the thyratron to exhibit a flat-top required for uniform amplification of the RF pulse.
But the ``flat top'' duration and quality of regulation are limited by the number of lumped elements in the PFN circuit.  
Additionally, the complex gain of the klystron is affected by the time-dependence of the high voltage applied through the PFN, so the tuning procedure must be carried out based on the shape of the RF pulse waveform post-amplification.  
Typically, only the amplitude of the RF pulse is optimized (i.e. flattened) with this tuning procedure, while realistically a phase shift of several degrees is still present over the output RF pulse duration.

Further optimization of the RF pulse amplitude and phase could be, and has been achieved by increasing the number of lumped elements in the PFN filter.  
Tuning, however, is still a cumbersome and dangerous procedure.  
The existing Mk3 PFN consists of eight sections of high voltage inductors and capacitors submerged in a large tank of dielectric oil and is unable to accommodate any further filter complexity within the present design.

Another approach to this regulation problem is the use of active feedback to provide a correction signal to either the high voltage pulse or input RF pulse.  
The most straightforward approach is to regulate the amplitude and phase of the input RF signal to compensate for any non-uniformity in the complex transfer function of the klystron.  
Such a feedback loop requires fast measurement of the amplitude and/or phase of the high power RF output and a means to feed this signal back to an amplitude/phase modulator on the RF source.  
Unfortunately, the long distances between components in the lab leads to latencies and long phase delays (hundreds of \si{\ns}).  
While the cable delays could in principle be reduced, ultimately the system latency is limited by the e-beam propagation delay in the klystron itself.
These effects make a high bandwidth and stable feedback loop impractical in this system.

Alternatively, given the good pulse to pulse repeatability of the high power RF system, and slow drift in the amplitude and phase perturbations observed in the output RF pulse, a compensation scheme can be applied.  
This compensation is not applied in real time to each pulse as in conventional feedback, but rather, is applied to subsequent pulses, which is termed feed-forward.  
The RF pulse amplitude and phase are measured in real time and the data are used to compute correction waveforms. 
Using arbitrary waveform generators, the corrections are applied to amplitude and phase modulators acting on the klystron input RF pulse.  
Subsequent measurements of the amplitude and phase are performed on the compensated high power RF pulse and allow iterative correction of any residual error.
Similar, but independent, feed-forward compensation systems have been implemented at several other accelerator facilities\cite{Zhang1993421,hartley1996aa,li1998aa,kawase2012aa}.  
Only recently, however, has direct quadrature demodulation been used to measure the amplitude and phase of the RF waveform to be compensated\cite{kasemir2005adaptive,hu2007rf}.

The present paper reports what we believe is the first application of a feed-forward system using the quadrature demodulation technique to the operation of an FEL and the diagnostics necessary to operate an inverse-Compton source.
Not only can the amplitude and phase of the RF pulse be normalized to remove the ripples observed in the RF and electron beam systems, but custom waveforms may be sculpted to optimize the performance of the free-electron laser and optical storage cavity-based x-ray source.
The measurement apparatus has also served as a versatile tool for diagnostics and analysis of signals related to the electron beam and RF power system.

This paper describes the design and performance of the feed-forward stabilization system developed and currently operated at UH.  
Section~\ref{sec:analytictech} gives an analytic description of the demodulation techniques used to construct the feed-forward error signal, while Section~\ref{sec:implementation} reviews the hardware and software used to implement this solution.  
Section~\ref{sec:results} describes the performance of the feed-forward phase compensation and the later addition of amplitude compensation capabilities.  
Finally, planned future improvements to the system are presented in Section~\ref{sec:future}.

\section{Analytic Technique}
\label{sec:analytictech}
\label{sec:directconversion}
The phase and amplitude of the RF waveform to be corrected are reconstructed using direct quadrature demodulation (IQ demodulation). 
The received RF signal is written as $V_\textsc{s}^\textsc{i}(t) = A(t) \cos(\omega t + \phi(t))$, where $A(t)$ and $\phi(t)$ are the amplitude and phase modulation components and $\omega = 2\pi(\RFG)$ is the RF carrier frequency.  
The signal is split, with one of the replicas delayed in phase by \ang{90}, giving $V_\textsc{s}^\textsc{q} = A(t) \sin(\omega t + \phi(t))$.

The two split signals are then separately mixed (multiplied) with an unmodulated reference signal $V_R = \cos(\omega t)$ at the RF frequency $\omega$ giving:

\begin{align}
I(t) = V_R V_\textsc{s}^\textsc{i}	&= A(t) \cos(\omega t) \cos(\omega t + \phi(t)) \notag \\ 
											&= \frac{A}{2} \cos\phi(t) + \frac{A}{2} \cos(2\omega t + \phi(t)) \\
Q(t) = V_R V_\textsc{s}^\textsc{q}	&= A(t) \cos(\omega t) \sin(\omega t + \phi(t)) \notag \\ 
											&= \frac{A}{2} \sin\phi(t) + \frac{A}{2} \sin(2\omega t + \phi(t))
\end{align}

A low-pass filter removes the second harmonic $(2 \omega)$, component yielding the in-phase and quadrature signals:

\begin{align}
	I(t) &= \frac{A}{2} \cos\phi(t) \\
	Q(t) &= \frac{A}{2} \sin\phi(t)
\end{align}

Now the amplitude and phase are reconstructed using:
\begin{eqnarray}
	A(t) = 2 \sqrt{I^2 + Q^2} \label{eqn:defamp}\\
	\phi(t) = \arctan (\frac{Q}{I}) \label{eqn:defphase}
\end{eqnarray}


The reconstructed amplitude and phase error signals from (\ref{eqn:defamp}) and (\ref{eqn:defphase}) are used to drive separate arbitrary waveform generators for the amplitude and phase feed-forward signals, respectively.
In the present compensation scheme, each correction signal is programmed to be directly proportional to the error signal and then shifted by a fixed time delay.
The proportionality constant and time delay are chosen empirically to minimize the fluctuations in subsequent pulses.

\section{Experimental Implementation}
\label{sec:implementation}
\subsection{Hardware}
\begin{figure*}
	\includegraphics[width=6in]{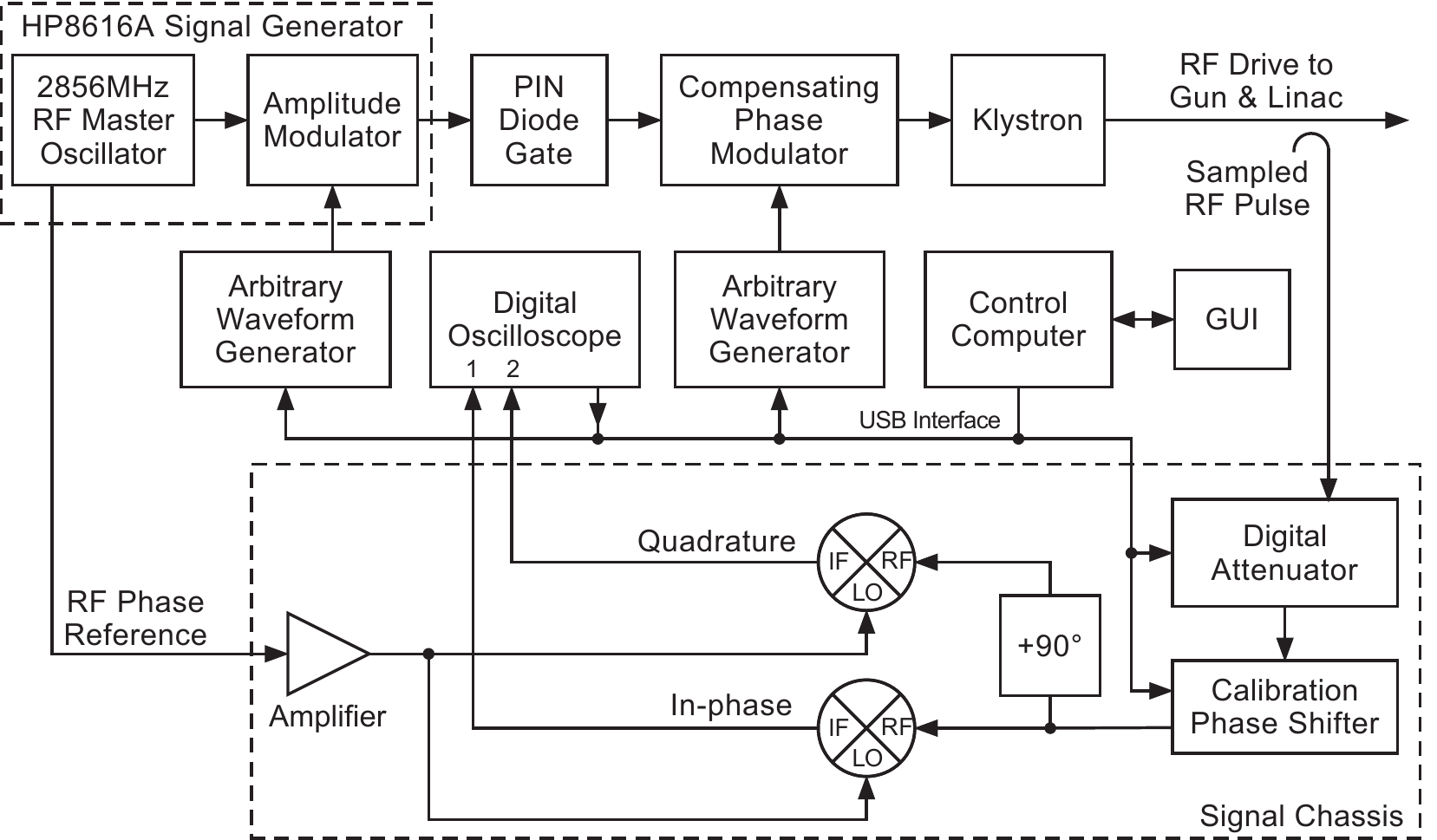}
	\caption{Block diagram of the MkV feed-forward system: The signal chassis demodulates RF pulses sampled from the high power linac drive.  A control computer acquires the pulse data from a digital oscilloscope and generates compensation waveforms.  Two arbitrary waveform generators apply the correction to the respective amplitude or phase modulator for all subsequent pulses.\label{fig:sysoverview}}
\end{figure*}

Figure~\ref{fig:sysoverview} shows the major components of the amplitude and phase compensation system.  
The RF signal chain for the MkV FEL consists of the components in the first row of the block diagram.  
An HP8616A klystron-based, microwave signal generator is used as the master oscillator for the entire RF system.  
The HP8616A includes a voltage controlled PIN diode attenuator that can be used to apply amplitude modulation from an external source.  
The CW signal is then chopped into \SI{4}{\us} pulses at \SI{4}{\Hz} using a PIN diode switch.
Next, a SLAC-designed voltage controlled phase modulator is used to apply a compensating phase modulation to the pulses.  
The modulator is capable of producing a \ang{180} phase shift with a \SI{10}{\volt} control signal.
The modulated RF pulses are finally amplified by the ITT klystron and the resulting $\sim\!\SI{30}{\mega\watt}$ pulses are delivered to the linac via an evacuated copper waveguide.
A directional coupler at the input to the linac is used to sample the RF pulses for feed-forward.  
The overall modulation bandwidth of this system is limited by the high power klystron to approximately \SI{3.5}{\MHz}.

The signal chassis contains the hardware necessary to demodulate the sampled RF pulses for analysis as described in Section~\ref{sec:directconversion}.
A Hittite Microwave digital attenuator provides up to \SI{31}{\dB} attenuation for calibration of input signal amplitude, while a second SLAC phase modulator is used to vary the absolute phase by up to \ang{180} for calibration purposes.
Both of these devices are controlled from a control PC connected via USB.

\begin{figure}
	\includegraphics[width=3.37in]{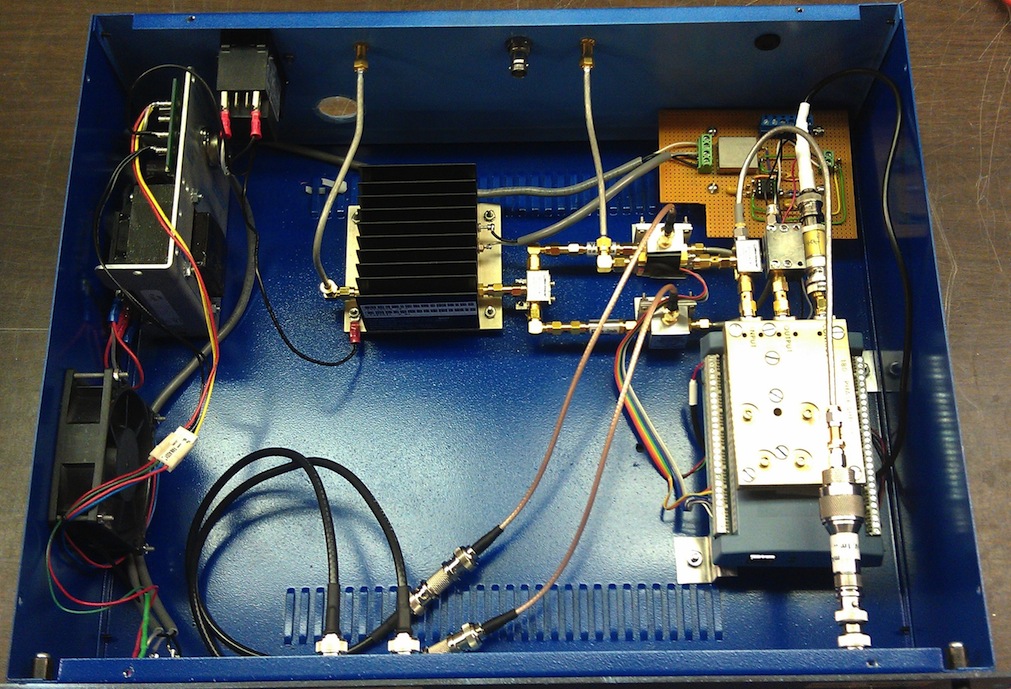}
	\caption{Feed-forward system hardware layout: The RF signal chassis contains the reference amplifier, mixers, digital attenuator, and phase shifter.  Power supplies and computer interface modules are also included.\label{fig:chassisphoto}}
\end{figure}

A Mini-Circuits ZX10Q-2-34-S+ hybrid splitter is used to divide the signal into two components of equal amplitude with a \ang{90} phase difference between them.  
Next, the split signals are separately mixed in MiniCircuits ZEM-4300+ double-balanced mixers with a phase reference to produce the in-phase and quadrature signals.  
The reference signal is derived from an unmodulated output of the HP8616A and divided with a \ang{0} power splitter.
This hardware is assembled in a 19~inch rackmount chassis (see Figure~\ref{fig:chassisphoto}) with associated power supplies and other standard components.

The pulse waveforms for the in-phase and quadrature signals are sampled using a Rigol DS1102E \SI{100}{\MHz}, 8-bit precision oscilloscope.  
The resulting 600 waveform samples at \SI{10}{\ns} intervals over \SI{6}{\us} are then transmitted over USB to a control PC running Linux.  
The control software, described in Section~\ref{sec:software}, derives compensation waveforms from the digitized in-phase and quadrature data.
Two Berkeley Nucleonics BNC645 14-bit, \SI{50}{\MHz} arbitrary waveform generators (AWGs) are programmed with the calculated compensation waveforms via USB, which are then applied to the amplitude and phase modulators for feed-forward to subsequent RF pulses.

The oscilloscope, AWG's, and PIN diode switch are all synchronized with a common trigger signal.

\subsection{Software}
\label{sec:software}
\begin{figure}
	\includegraphics[width=3.36in]{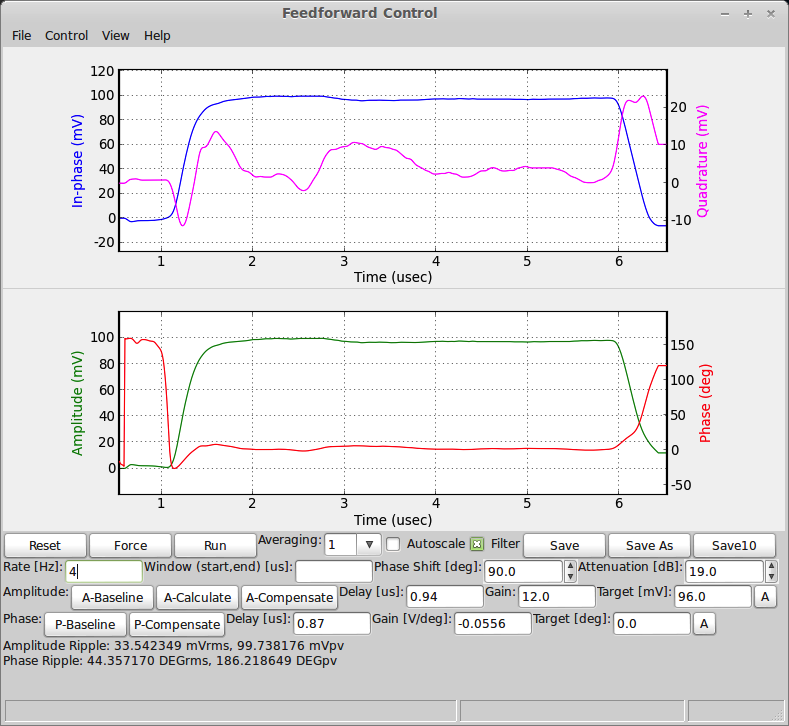}
	\caption{Feed-forward GUI: The graphical user interface (GUI) for the feed-foward system is implemented in wxPython\cite{wxPython} running on Linux.  The upper plot shows the raw sampled in-phase and quadrature signals, while the bottom plot shows the reconstructed amplitude and phase.  System commands are issued using the buttons at the bottom of the window.\label{fig:gui}}
\end{figure}

The feed-forward system is controlled from a PC running Linux in the FEL control room.  
The instrumentation is directly connected to the PC via USB cables.  
The control software is written in the Python programming language with a graphical user interface (GUI) utilizing the wxPython\cite{wxPython}, and Matplotlib\cite{matplotlib} toolkits.  
The layout of this GUI is shown in Figure~\ref{fig:gui}.
The oscilloscopes and arbitrary waveform generator are both compliant with the USBtmc interface standard, which allows for familiar GPIB style control via a modern USB connection.

The control software uses a modular design to facilitate future maintenance and upgrades.  
The GUI provides a user friendly subset of commands for day-to-day use of the system.   
A command line interface (CLI) provides more comprehensive access to all system functionality.  
The CLI allows scripts to automate common measurement and calibration tasks.  
These high-level interfaces are based on a low-level software module that provides programatic control over all functions of the system.  
Additional command line programs allow for offline analysis of captured data.


\section{Experimental Results}
\label{sec:results}
\subsection{Phase Compensation}
\label{sec:phasecomp}
The feed-forward system allows the compensation of the RF phase based on an RF measurement made at any point in the accelerator system.  
For example, the RF waveform sampled at the linac feed point may be used to derive a feed-forward compensation correcting the waveform at that point.  
Ultimately, it is the phase of the actual electron bunches that must be compensated, so the signal from one of several beam position monitors will eventually be sampled and corrected.  
Strip-line and image current beam position monitors (BPMs) provide a signal proportional to electron beam current at several positions along the beam-line and have sufficient bandwidth to resolve the \RFG bunch repetition rate of the beam current. 

\begin{figure}
	\includegraphics[width=3.375in]{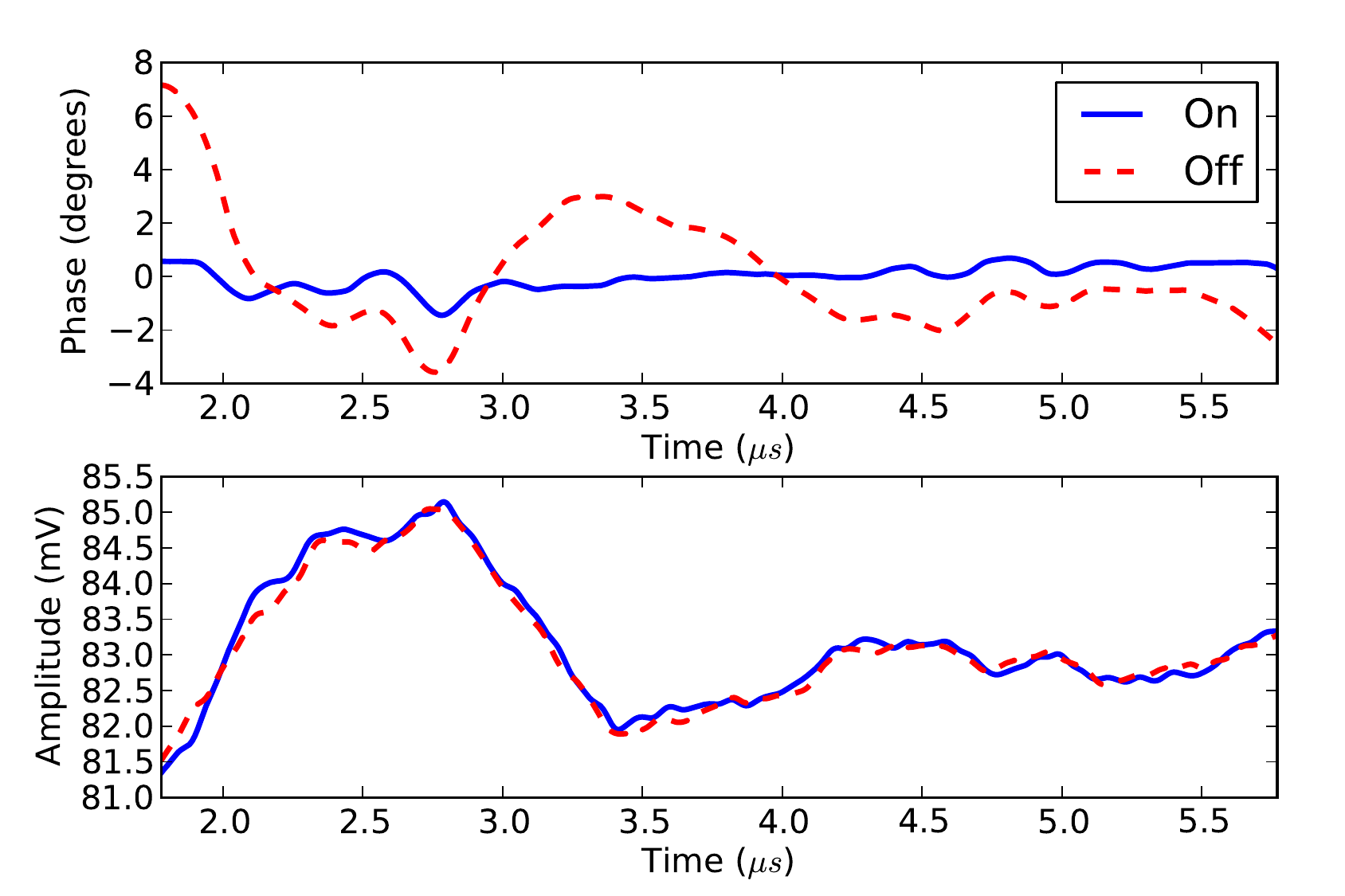}
	\caption{Effect of phase compensation: The application of a feed-forward phase compensation waveform results in an 80\% reduction in phase ripple.\label{fig:pmwaveform}}
\end{figure}

The system was initially commissioned with only phase compensation capability for RF pulses measured at the accelerator input port.  
The compensation waveform is simply a scaled and time shifted copy of the measured phase waveform.  
The appropriate delay was determined empirically by shifting the compensation waveform in \SI{10}{\ns} increments to minimize residual ripple in the measured phase. 
The phase ripple was also minimized for a scale factor corresponding to a negative unity gain through the system (a requirement of feed-forward systems).  
The empirically optimized scale factor corresponds to the measured gain of the SLAC phase shifter.  
Figure \ref{fig:pmwaveform} shows the effect of the phase compensation system on a \SI{4}{\us} window in flat-top region of the RF pulse waveform.  These data show that the phase ripple is reduced from $\ang{10.7} \textsc{pp} (\ang{2.1} \textsc{rms})$ to $\ang{2.1} \textsc{pp} (\ang{0.4} \textsc{rms})$ within the measurement window, an 80\% reduction.  
This measured phase ripple is limited by the $\ang{0.2} \textsc{rms}$ phase resolution imposed by the quantization error of the 8-bit oscilloscope used for waveform sampling.


This system has proven useful not only for feed-forward compensation, but also as a general purpose measurement tool for RF diagnostics.  
Beyond the linac drive system, the \RFM bunch repetition rate is present in every part of the system.  
The ability to easily measure RF amplitude and phase modulations yields a very useful diagnostic for understanding the micro-pulse structure of the electron beam and free-electron laser.  
The system has been successfully demonstrated with alternate RF sources such as beam position monitors and fast photodetectors.  
It may even be possible to use these signals for feed-forward compensation of the RF waveform to control the structure of these alternate source signals.

\subsection{Laser Performance}
\begin{figure*}
	\centering
	\includegraphics[width=6.41in]{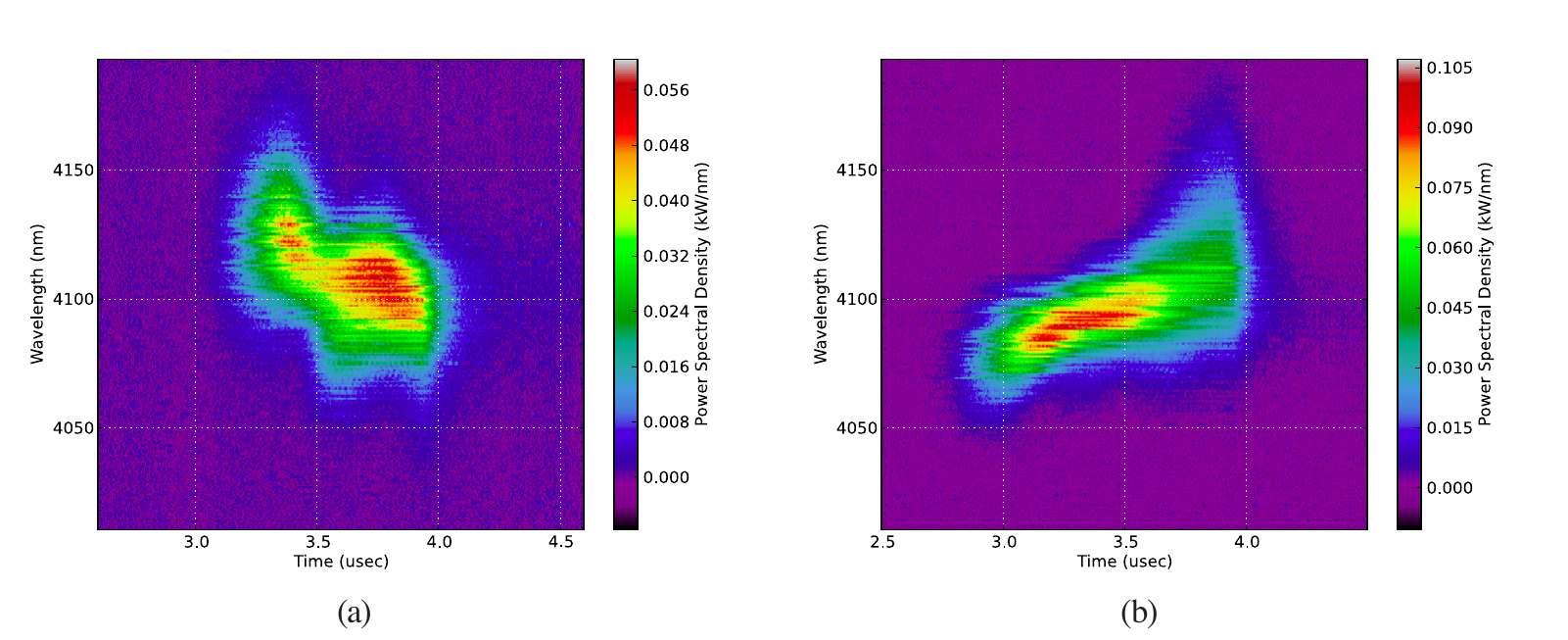}
	\caption{Laser spectrograms: Measured effect of feed-forward on laser performance (a) with phase compensation off and (b) with phase compensation enabled. \label{fig:spectrogram}}
\end{figure*}

\begin{figure*} 
	\centering
	\includegraphics[width=6.67in]{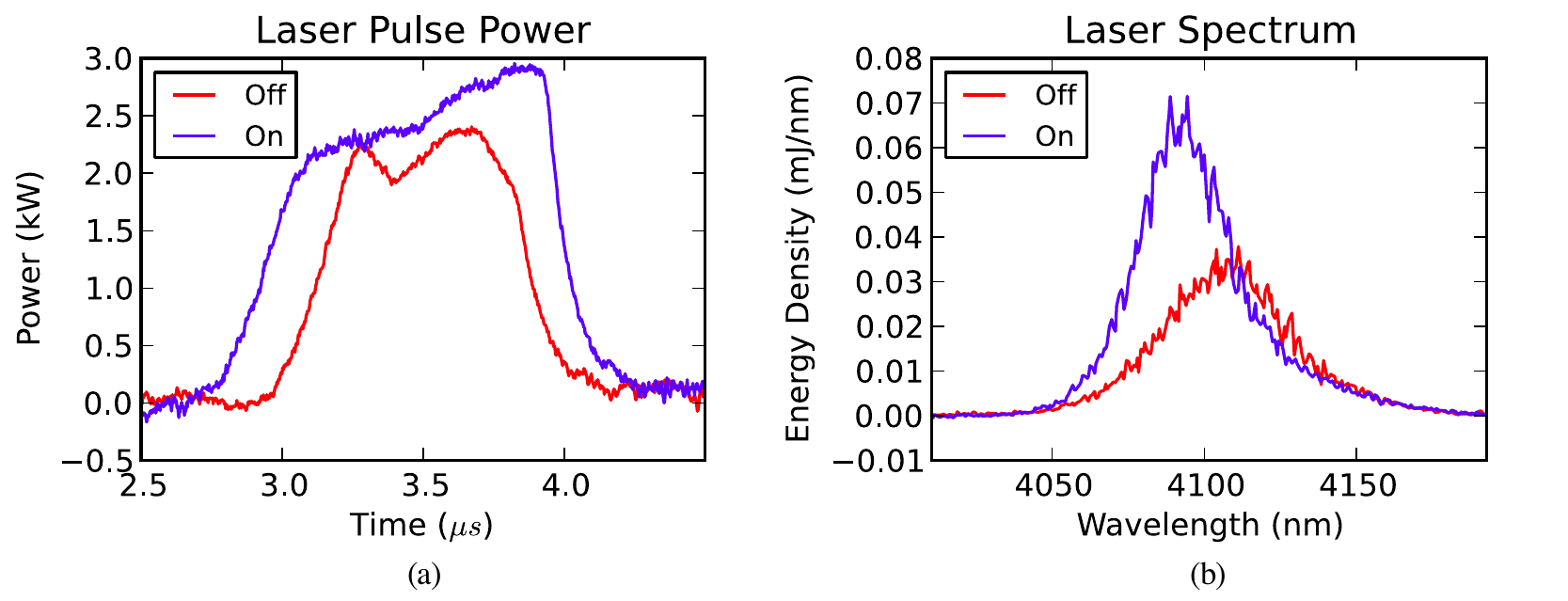}
	\caption{Measured effect of the feed-forward system on laser performance.  (a) The laser pulse power and (b) spectrum are computed as the integrals of the spectrogram plot in $\lambda$ and $t$ respectively.  The compensation lengthens the pulse duration by 40\%, while the spectrum narrows by 20\%.}
	\label{fig:comparepowerandspectrum}
\end{figure*}

Laser performance is evaluated using the spectrogram, or frequency vs. time plot, as a fundamental diagnostic of the performance of the free-electron laser.  
The spectrograms in Figure \ref{fig:spectrogram} show both the temporal and spectral evolution of FEL pulses.  
These data are the result of capturing the laser pulse waveform after filtering by a scanning monochromator.  
Since this measurement involves a slow spectral scan of the laser, the spectrogram comprises a record of the laser performance over many independent pulses.  
Random variations in the electron beam energy are evident from the horizontal bands seen in the image.
Figure~\ref{fig:comparepowerandspectrum} shows the projection of the two spectrograms onto their spectral and temporal axes.  These data clearly show a change in laser performance with phase compensation activated.
When compensation is applied the laser pulse turns on sooner, and the duration is extended by 67\%, while the FWHM spectral width sharpens by 20\%.

Without phase compensation (Figure~\ref{fig:spectrogram}a) there are two distinct peaks in the spectrogram, with the laser initially starting at a longer wavelength then lasing more strongly at a shorter wavelength.  
With phase compensation (Figure~\ref{fig:spectrogram}b), the instantaneous spectrum is substantially narrower and is pulled continuously towards longer wavelengths.  
This wavelength pulling agrees in both sign and magnitude with the frequency pulling phenomenon predicted by FEL theory\cite{benson1985tmf}, and is likely disrupted in Figure~\ref{fig:spectrogram}a due to the presence of phase perturbations in the electron beam.  
With our system, it should be possible to deliberately introduce a phase slew that compensates the frequency pulling effect, thus producing FEL macro-pulses better suited for use in the optical storage cavity-based x-ray source.  In addition, the technique of ``dynamic desynchronism''\cite{bakker1993dd} can be implemented using our system to improve laser turn-on time and stability.

To evaluate the intrinsic effectiveness of the compensation, we performed numerical simulations of the MkV FEL driven by electron pulse trains of constant energy subject to the phase perturbations depicted in Figure~\ref{fig:pmwaveform} and by electron pulses with no phase perturbation. 
These simulations are based on a one-dimensional numerical integration of the coupled Maxwell-Lorentz equations of motion describing the FEL interaction, in which the transverse profiles of the electron and optical beams are incorporated analytically\cite{crisafulli2001}. 
The present simulations were modified from earlier published studies by allowing the arrival time of the electron bunches to vary in accordance with the phase variations measured by the feed-forward system, which is an appropriate measure since the emission time of the electron bunches is primarily determined by the phase of the RF wave in the thermionic microwave gun. 
The remaining simulation parameters describing the electron beam, optical beam and FEL undulator were consistent with those of the MkV FEL\cite{Barnett1996mk3}.

\begin{figure}
	\includegraphics[width=3.375in]{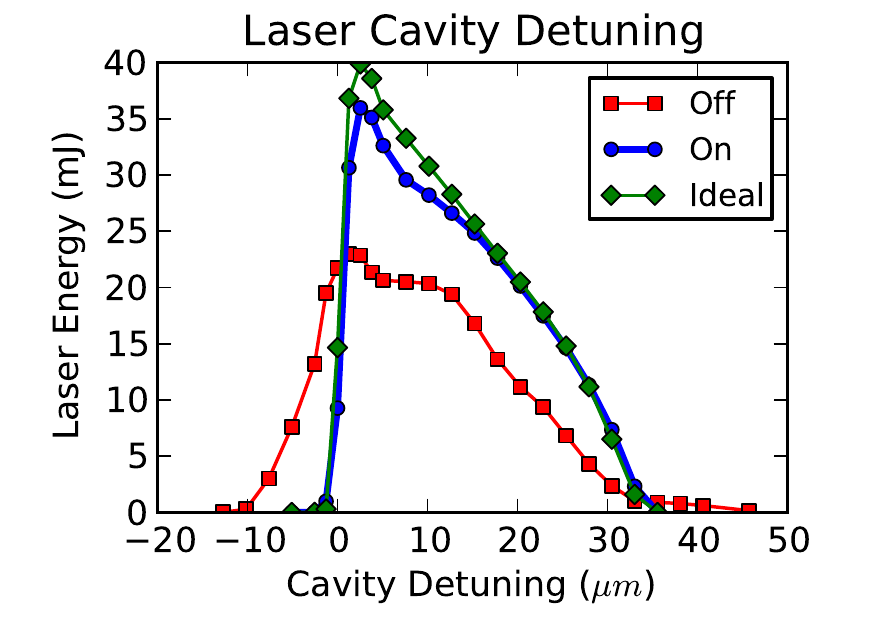}
	\caption{Simulated laser cavity detuning curves for uncompensated phase perturbation (boxes), compensated phase perturbation (circles), and no phase perturbation (diamonds).\label{fig:detuning}}
\end{figure}

The fundamental indicator of laser performance in an RF linac FEL is the cavity detuning curve of laser energy versus cavity length, which quantifies the mode locking efficiency of the picosecond laser pulses by the GHz-rate electron bunches. 
Figure~\ref{fig:detuning} displays the simulated cavity detuning curves corresponding to the uncompensated and compensated phase perturbations in Figure~\ref{fig:pmwaveform}, together with the detuning curve for an electron beam with no perturbation (perfect timing). 
The detuning curve for the uncompensated laser reveals significant degradation of laser energy compared to the phase-compensated detuning curve. 
The phase-compensated curve is more sharply peaked and displays the fundamental, asymmetric shape of the FEL detuning curve due to refractive effects in the electron beam\cite{colson1980free}, and yields essentially the same shape and width as the detuning curve for the perfect beam, with less than a 10\% reduction in pulse energy. 
These results indicate that the phase compensation achieved with the present system $(\ang{0.4} \textsc{rms})$ is sufficient to recover essentially optimum laser performance.
   
\begin{figure}
	\includegraphics[width=3.375in]{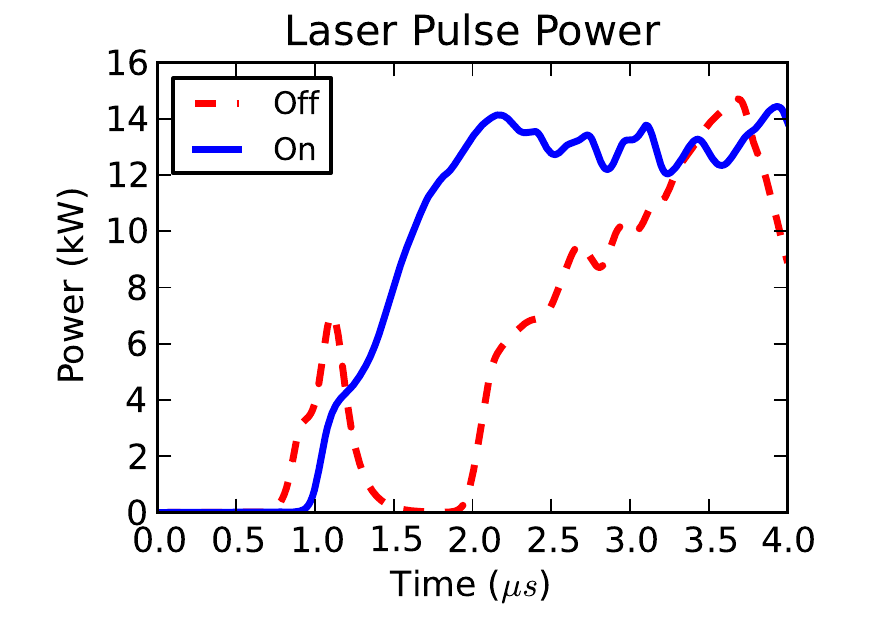}
	\caption{Simulated laser pulse power evolution with (solid) and without (dashed) electron beam phase compensation.\label{fig:simpower}}
\end{figure}

\begin{figure}
	\includegraphics[width=3.375in]{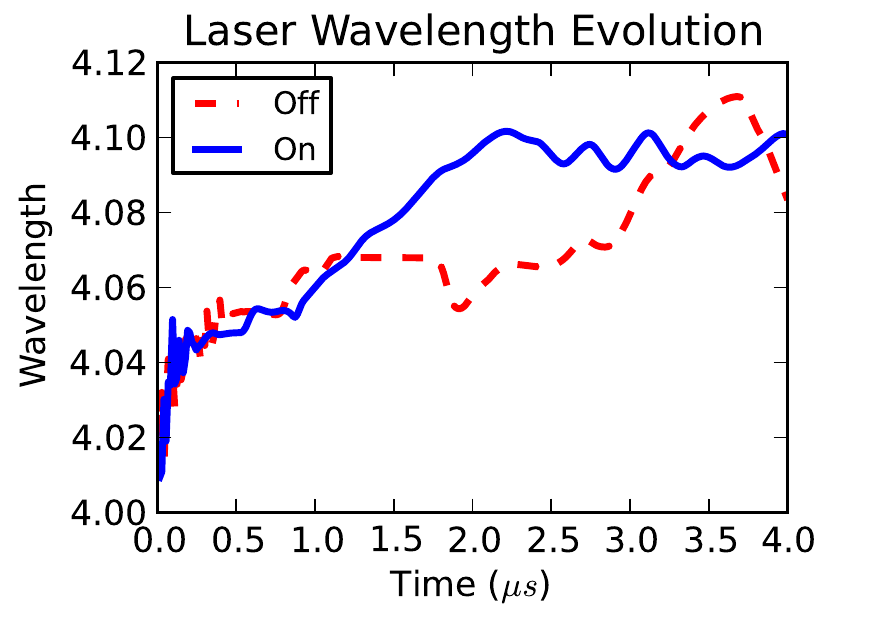}
	\caption{Simulated laser wavelength evolution with (solid) and without (dashed) electron beam phase compensation.\label{fig:simfreq}}
\end{figure}

To complete the numerical studies, we simulated the temporal evolution of the laser power and wavelength during the RF pulse for both the uncompensated and phase-compensated electron beams at the cavity lengths corresponding to maximum laser energy in Figure~\ref{fig:detuning}. 
The graph of laser power in Figure~\ref{fig:simpower} shows that the phase-compensated laser turns on substantially sooner than the uncompensated laser, as we observed empirically in Figure~\ref{fig:comparepowerandspectrum}a. 
The graph of laser wavelength in Figure~\ref{fig:simfreq} shows that frequency-pulling towards longer wavelengths, which is significantly disrupted in the uncompensated laser, is recovered in the phase-compensated laser, as we observed empirically in Figure~\ref{fig:spectrogram}b. These experimental and numerical results indicate that phase-compensation is an important capability for all future FEL operation. 

\subsection{Amplitude compensation}
While phase compensation has notably improved laser performance, significant ripple is still present in the amplitude driving the linac, causing an effective energy spread in the electron pulse train.
The subsequent addition of amplitude compensation capability has demonstrated a significant reduction in the energy spread of the electron beam, and should be able to correct e-beam perturbations introduced by beam loading in the electron gun, resulting in a more uniform electron beam current. 
Since the hardware was designed to capture the complex I/Q signal (see Section \ref{sec:directconversion}), the modifications necessary to include amplitude compensation involved upgrading the software to derive a feed-forward signal from the measured amplitude.  
An additional independent BNC Model 645 arbitrary waveform generator is programmed with the amplitude compensation waveform in the same manner as the phase compensation AWG.  The AWG output is then fed to the PIN diode amplitude modulator in the HP8616 microwave source.  
Additional controls added to the graphical user interface allow for optimization of the compensation delay and gain, and the target RF power level.

\begin{figure}
	\includegraphics[width=3.375in]{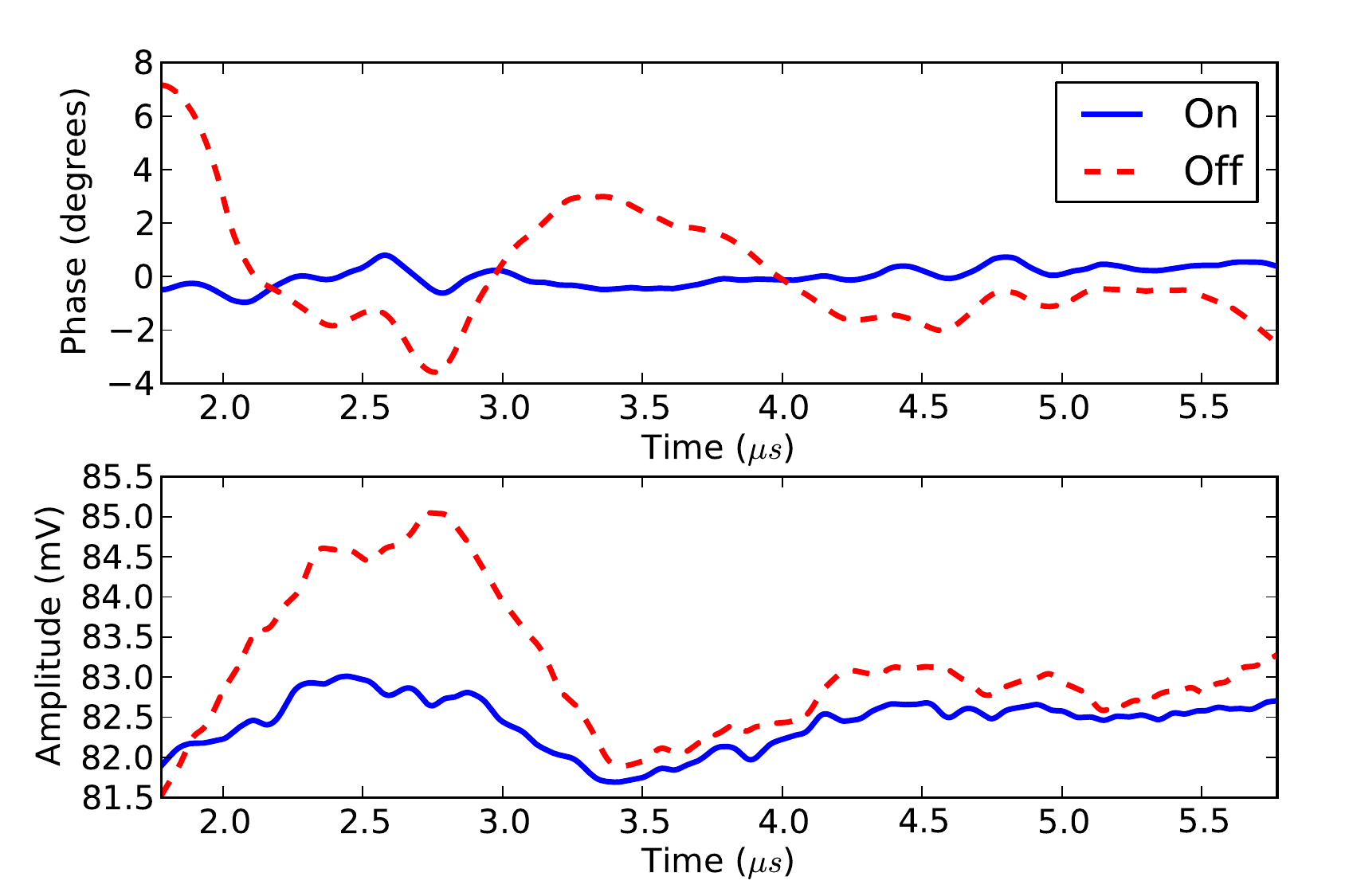}
	\caption{Effect of amplitude compensation:  Feed-forward phase and amplitude compensation result in an 85\% RF power ripple reduction. \label{fig:amresult}}
\end{figure}

Figure~\ref{fig:amresult} shows the RF waveform phase and amplitude measured at the linac input port with phase and amplitude compensation turned on and off.  
The phase ripple is reduced to \ang{0.4}, an improvement of 83\%, basically the same result as seen with phase compensation alone.
The amplitude ripple is now also reduced by 63\% corresponding to an 85\% reduction in RF power ripple.
The magnitude of both the residual amplitude and phase ripples are consistent with the theoretical performance limit of the system set by the use of an 8-bit oscilloscope.

\begin{figure}
	\includegraphics[width=3.375in]{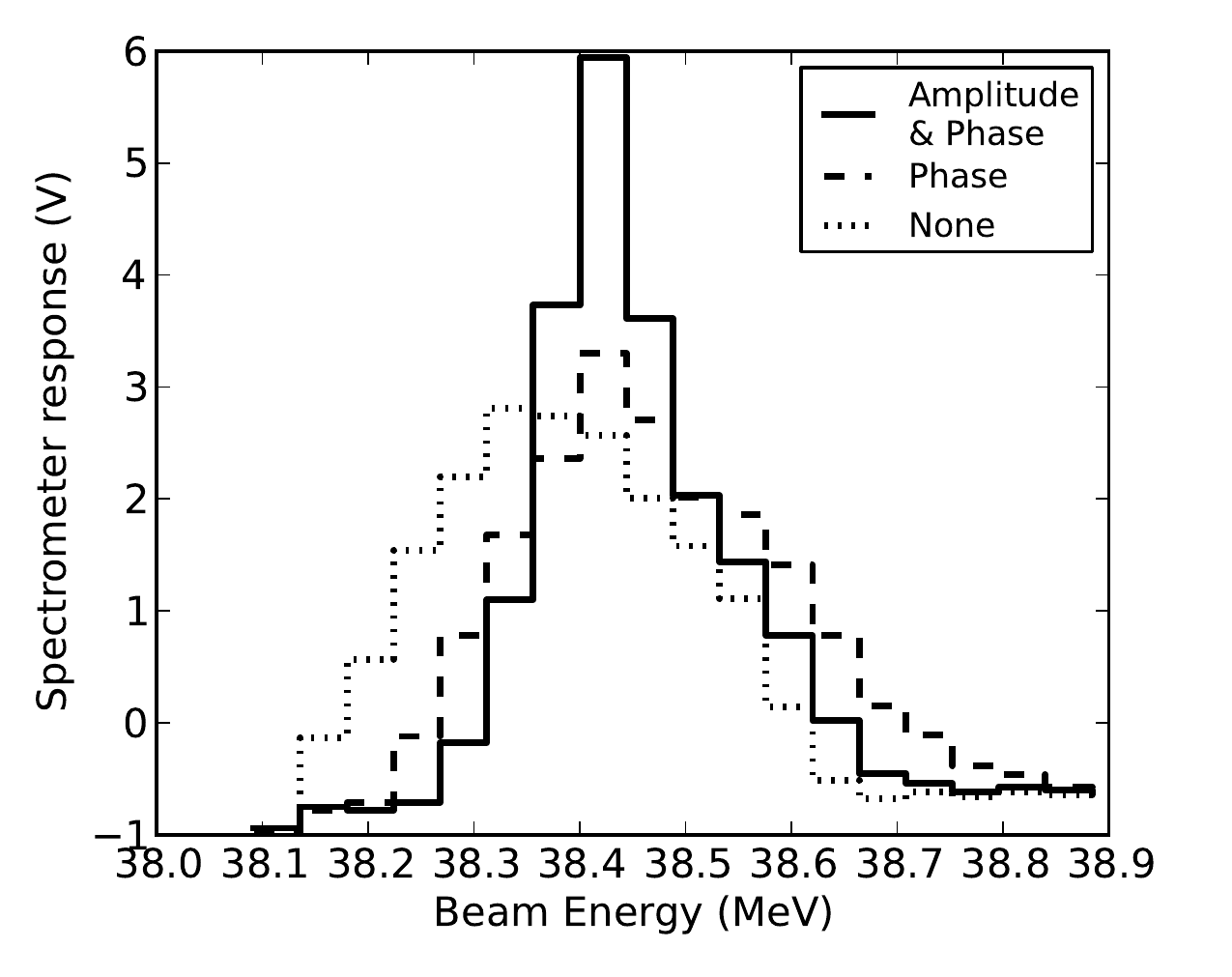}
	\caption{Measured electron beam energy spectrum with no compensation (dotted), phase compensation alone (dashed), and phase and amplitude compensation (solid).\label{fig:pamspectrum}}
\end{figure}

The resulting e-beam with phase and amplitude compensation exhibited greatly reduced energy spread.  Figure~\ref{fig:pamspectrum} compares the measured electron beam spectrum for amplitude and phase compensation, phase compensation only, and no compensation.
Phase compensation alone provides a 28\% reduction in the beam energy spread while both amplitude and phase decrease spectral width by 57\% to an energy spread of 0.3\% for a \SI{38}{MeV} electron beam.
Consistent with this reduction in energy spread, we also observed significantly improved focal spot performance at the mini-beta position of our e-beam chicane, which should greatly improve x-ray efficiency in our inverse Compton system.

\section{Future Developments}
\label{sec:future}
The preliminary results presented above have demonstrated that amplitude and phase compensation dramatically improve the quality of the electron beam and enhance laser performance of the MkV FEL at the University of Hawai`i.
A more sophisticated compensation algorithm should be capable of further reducing the residual phase and amplitude ripples in the RF signal.  
The present algorithm produces the compensation waveform by simply scaling and shifting the measured error waveform.  
This technique relies on the assumption that the feed-forward bandwidth is infinite (i.e. the impulse response $h(t)=\delta(t)$).  
By measuring the actual transfer function of the system, a more accurate compensation waveform can be constructed through deconvolution of the error waveform and the system impulse response.  

Proposed enhancements to the control software will allow for self-adaptive iterative compensation\cite{Zhang1993421} that we expect will reduce the effects of systematic errors due to nonlinearities in the feed-forward loop.  
While we have achieved good performance with a single iteration, this approach will also aid in the transition to a completely automated system.
The resolution of the measured amplitude, and thus the compensation effectiveness, are presently limited by the 8-bit dynamic range of the available digital oscilloscope. 
Performance could likely be improved by upgrading to higher resolution waveform sampling electronics.

Laser performance may be further enhanced through compensation based on measurements of the actual electron beam current and phase waveforms as mentioned in Section~\ref{sec:phasecomp}.  
While the amplitude and phase of the RF drive to the linac may be compensated with a simple linear correction model, the situation for the electron beam is more complex.  
The phase and energy spread of electron bunches exiting the linac are affected by the amplitude of the accelerating field, not only its phase.  
Implementation of a compensation scheme using the electron bunch phase measured by a strip-line BPM will require careful studies of the accelerator transfer function and the deconvolution of sampled beam data.
The hardware described in this paper is sufficient to meet the needs of such a system with the future work focused primarily on software development.

\section{Conclusion}
The feed-forward amplitude and phase compensation system recently implemented on the MkV FEL at the University of Hawai`i has proven very effective in reducing the amplitude and phase ripples on the accelerator RF drive and electron beam.  
Flat phase is essential to the efficient operation of the free-electron laser and an 80\% improvement has been realized.  
These performance enhancements are important steps to maximizing the peak power and pulse length of the FEL used to pump the storage cavity and will enhance the x-ray production rate in our inverse Compton system.  
The reduced energy spread of electron beam should also reduce the spectral width of the x-ray beam, further increasing the spectral brightness.
While a more sophisticated compensation algorithm may result in further reduction of the amplitude and phase ripple, the principal performance limitation is the use of 8-bit waveform sampling.  However, even with the dynamic range limited by the oscilloscope we have been able to achieve a phase ripple that simulations show is near optimal for laser operation.


%
%

%

\begin{acknowledgments}
Special thanks to Gary Varner for his advice and comments.  This work was funded under the Department of Homeland Security grant number 20120-DN-077-AR1045-02.
\end{acknowledgments}


%

\end{document}